\documentclass[%
reprint,showpacs,
superscriptaddress,
 amsmath,amssymb,
 aps,
]{revtex4-1}

\usepackage{multirow}
\usepackage{graphicx}
\usepackage{booktabs}
\usepackage{dcolumn}
\usepackage{amsmath}

\begin{document}

\title{A twelve-quadrupole correction for the interaction regions of high-energy accelerators}



\author{Javier  Fernando Cardona}
\email[]{jfcardona@unal.edu.co}
\affiliation{Universidad Nacional de Colombia, Bogot\'a, Colombia} 


\author{Yohany Rodr\'iguez  Garc\'ia}
\affiliation{Universidad Nacional de Colombia, Bogot\'a, Colombia}
 \affiliation{Universidad Antonio Nari\~no, Bogot\'a, Colombia}
\author{Rogelio Tom\'as Garc\'ia}
\affiliation{CERN,  Geneva CH-1211, Switzerland}

\date{\today}

\begin{abstract}
  Corrections of gradient errors in the interactions regions (IRs) of high energy colliders have  traditionally been made by changing  the strengths of quadrupoles that are common to both beams, such as the triplet quadrupoles. This article shows that  magnetic errors in the IR quadrupoles that are no common to both beams, such as the matching quadrupoles, can have an important influence and,  therefore, the correction should also include these quadrupoles. A correction based on twelve IR quadrupoles (common and no common) is presented and validated through MADX simulations. To estimate the strengths  of this correction, the action and phase in the inter-triplet space, the space that separates the two triplets of the IR, are required. A novel method to estimate these quantities is also presented. The main sources of uncertainties in this novel method are identified and compared to the current method that uses two beam position monitor within the inter-triplet space. Finally, LHC experimental data is used to
estimate the strengths of a twelve-quadrupole correction  in  the interaction region 1 of the LHC.  The resulting correction is compared with a six-quadrupole correction  estimated with another method called segment-by-segment (SBS).
\end{abstract}
\maketitle
\section{Introduction}
Gradient errors in the interaction regions (IRs) produce the largest
deviations in the optical model of a high-energy accelerator. The
correction of these errors is not only relevant to the overall performance of the
machine but also to ensure the best quality of the beam at the
interaction point (IP).

The ideal correction procedure is to measure the individual gradient errors of each IR magnet and change their strengths
to exactly compensate for each gradient error. However,
there is still no method to determine magnetic errors individually for each IR quadrupole. Current correction methods vary the strength of a few  IR quadrupoles in hopes of suppressing the effect of all  gradient errors present in
the IR. The first correction of this nature used in the LHC varied the
strengths of  two IR quadrupoles. These strengths  can be estimated
with two different methods, which are
segment-by-segment (SBS)~\cite{rtg_prst09,rtg_prst10,rtg_prst12,gv_phd}  and action and
phase jump analysis (APJ)~\cite{jfc_phd,jfc_prab17}, and both of them give
similar results. The two-quadrupole correction is
effective in suppressing the $\beta$ beating in the arcs. However, suppression of the $\beta$ beating in the IP is not guaranteed. To solve this problem, two different  corrections were  proposed.  The first is  a six-quadrupole correction with strengths that can
be estimated with SBS~\cite{lhc_2015} and the second is a four-quadrupole correction with strengths that can
be estimated with APJ~\cite{jfc_prab17}. It can be demonstrated that these two
corrections are also equivalent  and both of them effectively suppress the 
$\beta$-beating in the arcs and the IP as well. However, these
corrections work only  if the magnetic errors in the matching quadrupoles, the quadrupoles that are just outside the triplets, are small. Otherwise, a more general correction is required. In this paper, a twelve-quadrupole correction, which includes matching quadrupoles, is presented  and validated through MADX~\cite{mad} simulations and experimental data.

Estimates of corrector strengths in this paper are based on APJ and,
particularly, they depend on the action and phase in the
inter-triplet space, the space that separates the two triplets of the
IR. The current method to estimate these quantities do not have sufficient
accuracy to allow reliable estimates of the correction strengths. A
novel method to estimate action and phase in the inter-triplet space
with very low uncertainties is  presented in this paper.

The paper starts with  a  review of the APJ method in Sec.~\ref{apjsection}. Then, in Sec.~\ref{apjfromkmod}, the novel
method to estimate the action and
phase in the inter-triplet space is described. It is shown that
this new method has  significantly smaller uncertainties than the
uncertainties associated with the current method that uses two BPMs in the inter-triplet space.
This new method uses  \textit{k}-modulation  measurements and the action and phase
that are independent  of the longitudinal position $s$: the action and
phase constants. Sec.~\ref{rotation}  describes how these constants
can be  measured accurately. Applying this new  development to LHC
experimental data,  the strengths of a correction  that uses only
common quadrupoles are estimated in Sec.~\ref{appen}. Comparisons
between the strengths obtained from beam 1 data and beam 2 data suggest
that magnetic errors in the no common quadrupoles are significant,
which leads to the more general twelve-quadrupole  correction
mentioned earlier. This correction  is introduced  and tested with
simulations in Sec.~\ref{matching}. Finally, the strengths of a
twelve-quadrupole correction are estimated from experimental LHC turn-by-turn (TBT) data
and compared  with a  six-quadrupole correction  estimated with SBS in  Sec.~\ref{estimates}.

\section{The Action and Phase  Jump Method}\label{apjsection}
It has been shown in~\cite{jfc_prst09, jfc_prab17}  that the APJ method allows the mathematical description of a one-turn particle
trajectory in the presence of linear magnetic errors with \par
\begin{equation}
z(s) = \sqrt{2 J(s) \beta_n(s)} \sin[\psi_n(s) - \delta(s)],
\label{apj_betatron}
\end{equation}
where $z(s)$ is the particle transverse position (either $x$ or $y$) with
respect to the closed orbit,  $\beta_n(s)$ and $\psi_n(s)$ are the
nominal lattice functions, and $J(s)$  and $\delta(s)$ are the actions and phases that, unlike the action and phase of the conventional betatron equation, jump at magnetic error locations.

These jumps allow to estimate the  deflection $\theta$, also called
magnetic kick, that a
particular magnetic  error  produces in the particle trajectory with \par
\begin{equation}
|{\theta}| = \sqrt{\frac{2 J_0 + 2 J_1 + 4 J_0  J_1 \cos(\delta_1 -
    \delta_0)}{\beta_n(s_{e})}},\label{tetaexp}
\end{equation}
where $J_0$, $J_1$, $\delta_0$, and $\delta_1$ correspond to the
actions and phases immediately to the left and  to the right of
 $s_e$, the axial location of the magnetic error. Assuming that the
magnetic error has only quadrupole components, the
following relationships are valid 
\begin{subequations}
 \label{teta1mag} 
\begin{eqnarray}
\theta_x & = &-B_1 x_e + A_1 y_e, \\
\theta_y & = & B_1 y_e + A_1 x_e, 
\end{eqnarray} 
\end{subequations}
where $B_1$ and $A_1$ are quantities proportional to the normal and
the skew quadrupole components of the magnetic error that caused the
deflection of the particle trajectory, and $x_e$  and $y_e$ correspond
to the  position of the particle evaluated at $s_e$. Since the deflections in both
planes can be estimated with Eq.~(\ref{tetaexp}), it is also possible
to estimate the numerical  values of $B_1$ and $A_1$ using  Eq.~(\ref{teta1mag}).

In practice, $J$ and $\delta$ at a particular location in the
accelerator are estimated using the trajectory measurements of two
adjacent BPMs, $i$ and $i+1$, as follows:
\begin{eqnarray}
J_{i+1} = && \frac{ \left (z_i/\sqrt{{\beta_n}_i} \right )^2 + \left
    (z_{i+1}/\sqrt{{\beta_n}_{i+1}}
  \right )^2}{ 2 \sin^2({\psi_n}_{i+1} - {\psi_n}_i)} \nonumber \\
          &&  - \frac{ z_i z_{i+1} \cos({\psi_n}_{i+1} -
  {\psi_n}_i)}{\sqrt{{\beta_n}_i{\beta_n}_{i+1}}   \sin^2({\psi_n}_{i+1} - {\psi_n}_i)},
\label{jandpsia}
\end{eqnarray}
and 
\begin{equation}
 \tan \delta_{i+1} = \frac{ (z_i/\sqrt{{\beta_n}_i}) \sin {\psi_n}_{i+1}-
  (z_{i+1}/\sqrt{{\beta_n}_{i+1}}) \sin {\psi_n}_i}{ (z_i/\sqrt{{\beta_n}_i}) \cos
  {\psi_z}_{i+1}-(z_{i+1}/\sqrt{{\beta_n}_{i+1}})\cos {\psi_n}_i}.
\label{jandpsib}
\end{equation}
This process is repeated for all  adjacent BPM pairs in the
accelerator, which makes possible to find  $J$ and $\delta$ as
function of $s$. Because the number of BPMs is limited,  it is not possible
to estimate the actions and phases associated with every accelerator
magnet. Only  the actions and phases associated with certain group of
magnets can be estimated. In the LHC, for example, the actions and phases
immediately to the left and right  of the high luminosity IRs
can be easily identified, as seen in Fig.~1 of~\cite{jfc_prab17}.  Although these actions
and phases are not sufficient to estimate the deflections produced by
every individual error in the magnets within a particular IR,  it is possible
to estimate an equivalent magnetic kick for the entire IR using
Eq.~(\ref{tetaexp}). Similar to the  kick of a magnetic error, the
equivalent  kick can also be expressed based on  its
magnetic quadrupole components  ${ B_1}_{x,e}$, ${ B_1}_{y,e}$,  and
$A_{1,e}$, as follows:
 \begin{subequations}
 \label{tetaIRmag} 
\begin{eqnarray}
\theta_{x,e} & = & -{B_1}_{x,e} x_e + A_{1,e} y_e,\\
\theta_{y,e} & = &  {B_1}_{y,e} y_e + A_{1,e} x_e.
\end{eqnarray} 
\end{subequations}
If Eq.~(\ref{tetaIRmag}) is used with two one-turn beam trajectories,
it is possible to estimate the quadrupole
components of the equivalent kick  as 
\begin{subequations}
\label{equierror}
\begin{eqnarray}
{ B_1}_{x,e}  &= & \frac{y_{e1} {\theta_{x_2,e}} - y_{e2} {\theta_{x_1,e}}   } {x_{e1} {y_{e2}} -{x_{e2}} {y_{e1}}}, \\
{ B_1}_{y,e} &= & \frac{x_{e1}  {\theta_{y_2,e}} - x_{e2}
                  {\theta_{y_1,e}}   } {x_{e1} {y_{e2}}
                  -{x_{e2}} {y_{e1}}},\\
A_{1,e} &= & \frac{x_{e1} {\theta_{x_2,e}} - x_{e2}
             {\theta_{x_1,e}}   } {x_{e1} {y_{e2}} -{x_{e2}}
             {y_{e1}}}, 
\end{eqnarray}
\end{subequations}
 where the numerical subscripts are used to differentiate variables
 that belong to one trajectory or the other. 

The quadrupole components of the equivalent kick can be used to
estimate a correction that suppresses the effect of  all magnetic
errors in the IR. For the normal quadrupole errors, this suppression is
achieved by changing the strength of two out of the six normal IR quadrupoles  so that
that the equivalent kick generated by these strengths has quadrupole components
$B^{(c)}_{1x}$ and $B^{(c)}_{1y}$  that are equal but opposite to the
quadrupole components of the original equivalent
kick. This leads to \par
\begin{equation}
 \label{expcorrquads} 
 { B_1}_{z,e} =- B^{(c)}_{1z},   
 \end{equation}
and
\begin{subequations}
 \label{intlincomp} 
\begin{eqnarray}
 { B_1}_{x,e}&=&-\frac{ \Delta {K_1}_a I_{x,a}  + \Delta {K_1}_b  I_{x,b}}{\beta_{n,x}(s_e)}
, \label{intlincompa}    \\
 { B_1}_{y,e}&=&-\frac{\Delta {K_1}_a I_{y,a}  + \Delta {K_1}_b  I_{y,b} }{\beta_{n,y}(s_e)}   \label{intlincompb}, 
\end{eqnarray}
\end{subequations}
where the measured quadrupole components of the equivalent kick have
been related to the strength changes  $\Delta {K_1}_a$ and $\Delta {K_1}_b$   needed in quadrupoles $a$ and
$b$ to suppress the effect of the normal
quadrupole errors present in the IR. Also in these equations,
$\beta_{n,z}$ (with $z$ representing either the $x$ or the $y$ plane) represent the nominal $\beta$ functions, and  $I_{z,i}$  corresponds to the integrals defined by \par
\begin{equation}
\label{originalint}
I_{z,i} =\int^{s_{ri}}_{s_{li}} ds' \beta_{n,z}(s'),
\end{equation}
where $s_{li}$ and $s_{ri}$ are the longitudinal positions of the left and 
right sides of  magnet $i$, which can be either $a$ or $b$.

The strengths  $\Delta {K_1}_a$ and $\Delta {K_1}_b$ can be
found by inverting Eq.~(\ref{intlincomp}) resulting in
\begin{subequations}
\label{stcorr}
 \begin{eqnarray}
\hspace{-0.5cm} \Delta {K_1}_a&=&\frac{ 
 { B_1}_{y,e} \beta_{n,y}(s_e) I_{x,b} - { B_1}_{x,e} \beta_{n,x}(s_e)
                                  I_{y,b}   } {
I_{x,a} I_{y,b}   -  I_{x,b} I_{y,a} },\label{stcorra} \\
\hspace{-0.5cm}\Delta {K_1}_b&=& \frac{ { B_1}_{x,e} \beta_{n,x}(s_e) I_{y,a} - { B_1}_{y,e} \beta_{n,y}(s_e)  I_{x,a}} {I_{x,a} I_{y,b}   -  I_{x,b} I_{y,a}}. \label{stcorrb}
\end{eqnarray}
\end{subequations}
One of the  effects produced by normal quadrupole
errors present in a particular IR is $\beta$-beating. When only two
IR quadrupoles are used for correction, with  strengths
estimated  with Eq.~(\ref{stcorr}), the  $\beta$-beating is
effectively suppressed everywhere in the ring.  The suppression,
however, is not complete; a significant $\beta$-beating can still
remain in the IP as shown with the red curve in Fig.~7 of~\cite{jfc_prab17} . 

To solve this problem, the IR magnets can be divided in two groups:
the left triplet magnets and the right triplet magnets. If the action
and phase in the inter-triplet space   can be known, the
equivalent kicks corresponding to each triplet can be
estimated. Therefore,  two  strengths can be estimated for each
triplet with Eq.~(\ref{stcorr}). This leads to a  correction
with four quadrupole per IR instead of two per IR, which effectively
suppress  the  $\beta$-beating everywhere in the ring including the IP
as shown with the blue curve  in Fig.~7 of~\cite{jfc_prab17}.
\section{Action and Phase in the inter-triplet
  space}\label{apjfromkmod}
The action and phase in the inter-triplet space is currently 
obtained with the two BPMs closest to the IP (BPMSWs) using
Eqs.~(\ref{jandpsia}) and~(\ref{jandpsib}). This method  does not have
sufficient accuracy to allow reliable estimates of the correction strengths. A novel method to estimate these quantities is presented in this section.

\begin{figure}[h]
\centering
\includegraphics{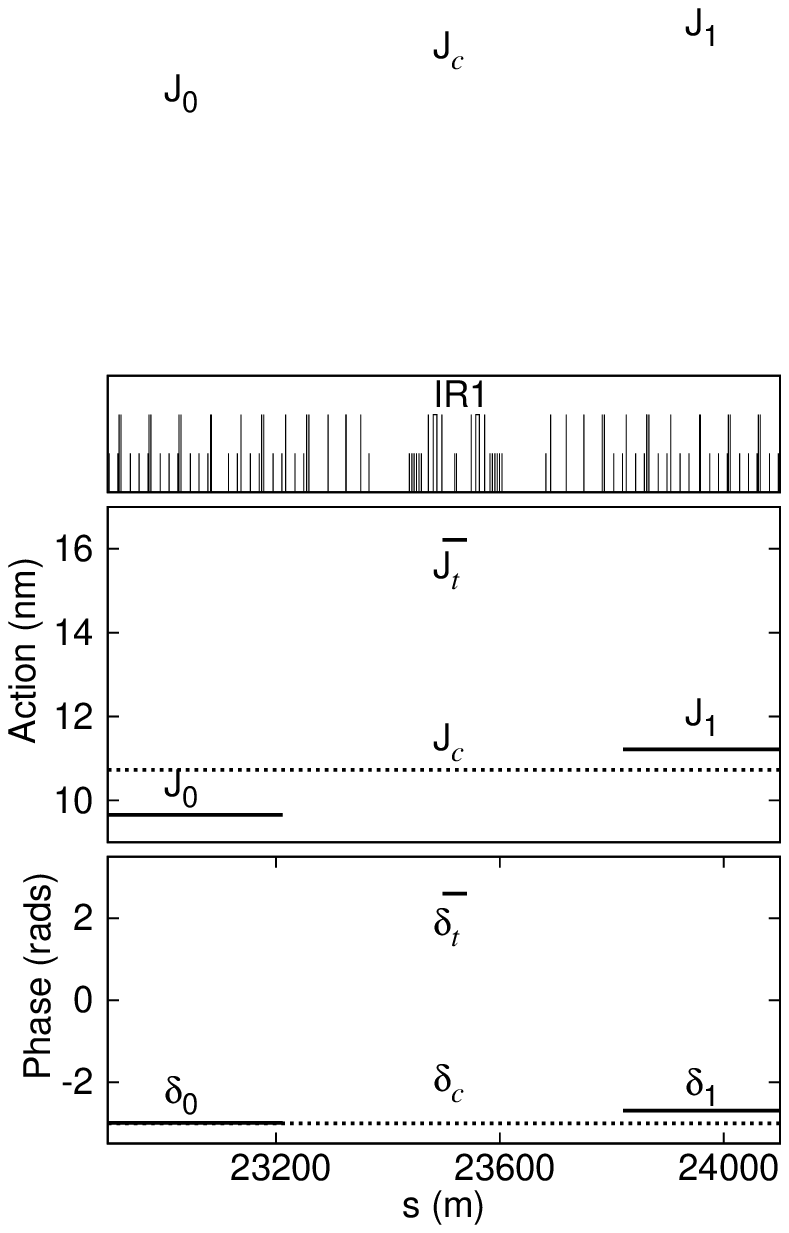}
\caption{\label{APJvsReal}  The action and phase variables  used in APJ
  analysis (solid lines) are illustrated, as well as the  action and phase
  constants (dotted lines). All the actions and phases were obtained from a simulated    
  average trajectory  of the LHC (beam 1).   In the
  upper part of the figure, the optical lattice is represented by
  long and short  vertical bars corresponding to quadrupole and dipole
magnets respectively. The two triplets of IR1 and  the inter-triplet
space can be seen just below the ``IR1'' label.}
\end{figure}

 Assume that  IR1 of the LHC  is been configured as a high luminosity IR
with a $\beta^* =$ 40 $cm$ ($\beta^*$ is the value of the
$\beta$-function  at the IP). Also assume that  magnetic quadrupole errors are
present only in the IR. If a one-turn particle trajectory is generated
with this LHC lattice, the corresponding APJ description, derived from Eq.~(\ref{apj_betatron}), is\par
\begin{equation}
z(s) = \begin{cases}
\sqrt{2J_0 \beta_n(s)} \sin{[\psi_n(s) - \delta_0]} &
\textnormal{arc  left of IR1} \\
\sqrt{2J_t \beta_n(s)} \sin{[\psi_n(s) - \delta_t]} & \textnormal{inter-triplet space}   \label{apjeqb}\\
\sqrt{2J_1 \beta_n(s)} \sin{[\psi_n(s) - \delta_1]} &     \textnormal{arc right of IR1}
\end{cases}
\end{equation}
where $\beta_n(s)$ and $\psi_n(s)$ correspond to the nominal lattice
functions while  $J$ and $\delta$ correspond to the actions and phases
used in APJ analysis. The subscripts $0$ and $1$ are used to label
variables in the arc  that are to the left and to the right of IR1
respectively. The subscript  $t$ is used to label variables corresponding to  the inter-triplet space (see also Fig.~\ref{APJvsReal}).

It  is also possible to use the conventional betatron equation  to mathematically
describe the same one-turn particle trajectory as follows \par
\begin{equation}
z(s) = \sqrt{2 J_c \beta_r(s)} \sin\left[ \psi_r(s) -\delta_c \right],
\label{betatron}
\end{equation}
where $\beta_r(s)$ and  $\psi_r(s)$ are the lattice functions that include
magnetic errors  and $J_c$ and $\delta_c$ are the action and phase constants.\par

In the inter-triplet space  Eqs.~(\ref{apjeqb})  and~(\ref{betatron})  lead to
\begin{eqnarray}
z(s) & =&  \nonumber\\
      & = & \sqrt{2J_c \beta_r(s)} \sin{[\psi_r(s) - \delta_c]} \label{RandIT}\\
     &=& \sqrt{2J_t \beta_n(s)} \sin{[\psi_n(s) - \delta_t]} \nonumber.
\end{eqnarray}
The beta and phase functions in the inter-triplet space are given by well-known analytical formulas.  These formulas depend on the minimum value of beta function, which is usually denoted by the symbol $\beta_w$, and the difference between the axial location of  $\beta_w$ and the axial location of the IP, which is generally known as the waist shift $w$. If these formulas are used in Eq.~(\ref{RandIT}), the following relationships for the action and phase in the inter-triplet space can be deduced

\begin{equation}
J_t = J_c \frac{\beta_{w_n}}{{\beta_w}_r} {\cos^2 \gamma_c}  \left( 1 +
  \tan^2 \gamma_{t} \right),
\label{JIT}
\end{equation} 
and 
\begin{equation}
\delta_t = \psi_n(s_{t}) + \arctan{  \left(\frac{L +w_n}{\beta_{w_n}}
  \right)} - \gamma_{t},
\label{deltaIT}
\end{equation}
where 
\begin{equation}
\tan \gamma_{t} = \frac{w_n - w_r +  {\beta_w}_r \tan\gamma_c}{\beta_{w_n}}, 
\label{tanIT}
\end{equation}
and
\begin{equation}
\gamma_c =\psi_r(s_{t}) + \arctan {  \left(\frac{L + {w}_r}{{\beta_w}_r} \right)} - \delta_c.\label{gammaR}
\end{equation}
In these expressions,  the subscripts $n$ and $r$ are used to distinguish the nominal variables
from  variables associated to the lattice with errors, $s_t$
corresponds to the axial location where the inter-triplet space starts,
and $L$ corresponds to  half the length of the inter-triplet space.

Expressions~(\ref{JIT}) to~(\ref{gammaR})  depend on experimental variables
that  are  routinely obtained  in the LHC  [$\psi_r(s_{t})$,
$w_r$, and ${\beta_w}_r$] and variables that can be obtained directly  from the
nominal model of the accelerator [$\psi_n(s_{t})$, $w_n$,
${\beta_w}_n$, and $L$]. In addition, they depend on  the action and
phase constants, which can be obtained from the experimental TBT data
sets as shown in Sec.~\ref{rotation}.

The three experimental variables required to estimate $J_t$ and
$\delta_t$ are obtained using  two different techniques. To obtain
${\beta_w}_r$ and $w_r$, a technique based on
\textit{k}-modulation is used~\cite{kmod}.  The general idea of this
technique is to change the strength of the two quadrupoles closest to
the IP and record the corresponding changes of the betatron tunes
in both planes. From this data, very accurate estimates of  the average $\beta$
functions of the two quadrupoles involved can be obtained. The values
of  ${\beta_w}_r$ and $w_r$ are  obtained later through analytical
equations that relate these variables with  the average $\beta$
functions. 

To obtain  the lattice functions with errors  [$\beta_r(s)$ and $\psi_r(s)$], a technique based on Fourier
analysis of TBT data is used~\cite{castro}. In this
technique the $\psi_r(s)$ functions are obtained directly  from the phase
resulting from the Fourier analysis in each BPM data set, while the
$\beta_r(s)$ functions are obtained through equations that relate the
phase advances between three consecutive BPMs and their nominal $\beta$
functions.
 
The action and phase in the inter-triplet space 
can also be obtained using  the two BPMs closest to the IP. Therefore, it is possible to test Eqs.~(\ref{JIT}) and~(\ref{deltaIT})   comparing the results of both methods.  For this
purpose  simulated TBT data  is   generated with  MADX for a LHC lattice with
quadrupole errors in IR1. In this simulation the ``experimental
values''   ${\beta_w}_r$, $w_r$, and   $\psi_r(s_{t})$   are   obtained directly from  Twiss files generated by MADX
for the lattice with errors while $J_c$ and $\delta_c$ are obtained
from the simulated TBT data.  The four kinds of average max
trajectories defined in~\cite{jfc_prab17} (see also Sec.~\ref{estimates}) are obtained from the simulated TBT data, and  $J_t$  and $\delta_t$ are obtained for every
trajectory using  Eqs.~(\ref{JIT}) and~(\ref{deltaIT}) and also using
the BPMSWs. In all cases there is an agreement of seven significant
figures between the two methods for both  quantities.

Since the uncertainties of all the experimental variables in Eqs.~(\ref{JIT}) and~(\ref{deltaIT})   are known (see
Table~\ref{sources}), the propagated uncertainties  $\Delta J_t$ and
$\Delta \delta_t$ can be estimated.  These uncertainties were estimated
for the four average trajectories generated in the previous simulation. The maximum values are
shown in the first row of  Table~\ref{incerti}.

\begin{table}[h]
\caption{\label{sources}  Uncertainties of the experimental
  variables required to estimate $J_t$ and $\delta_t$. The sources
  where these uncertainties were extracted are also listed.}
\begin{ruledtabular}
\begin{tabular}{l l  c}
\multicolumn{1}{c}{Exp. Variable}
 &
\multicolumn{1}{c}{Uncertainty}
&
\multicolumn{1}{c}{Extracted from:}\\
\colrule
$\qquad \psi_r(s_t)$ &6 mrads & \cite{psk_ipac16}  \\
$\qquad w_r$ & 1 cm &  \textit{k}-modulation experiments\\
$\qquad J_c$   & 0.5 \% & Sec.~\ref{rotation} \\
$\qquad \delta_c$  &  2 mrads  & Sec.~\ref{rotation}\\
$\qquad \beta_{w_r}$ & 0.3 mm &  \textit{k}-modulation experiments\\
\end{tabular}
\end{ruledtabular}
\end{table}

\begin{table}[h]
\caption{\label{incerti} Uncertainties associated with the estimates of action and
  phase in the inter-triplet space due to  the uncertainties in
  Table~\ref{sources}. These uncertainties are compared to the uncertainties of the method that uses  2 BPMSWs  with a gain error of 1\% in one of the BPMs.}
\begin{ruledtabular}
\begin{tabular}{l D{.}{.}{-1}  c}
 &
\multicolumn{1}{c}{$\Delta J_t$}
&
\multicolumn{1}{c}{$\Delta \delta_t$}\\
\multicolumn{1}{c}{Method}
&
\multicolumn{1}{c}{ (\%)}
 &
\multicolumn{1}{c}{ (rads)}\\
\colrule
Eq.~(\ref{JIT}) and Eq.~(\ref {deltaIT}) &2.7 & 0.015  \\
BPMSWs  &32.0 & 0.164 \\
\end{tabular}
\end{ruledtabular}
\end{table}
For comparison purposes the uncertainties associated with the method
using two BPMSWs are estimated. In  this method BPM gain errors are
the most important sources of uncertainty. Even  assuming the best
BPM calibration  achieved in the LHC so far (1\%  gain error),  the
corresponding $\Delta J_t$ and $\Delta \delta_t$ (second row of Table~\ref{incerti})  are significantly larger than the
uncertainties associated with Eq.~(\ref{JIT})  and Eq.~(\ref{deltaIT}). 

\section{Estimating  the action and phase
  constants}\label{rotation}
If one turn trajectories are well described by
Eq.~(\ref{betatron}),  $J_c$ and $\delta_c$  can be
estimated using, for example,  Eqs.~(\ref{jandpsia}) and~(\ref{jandpsib})  with only one pair of BPMs. However,
three sources  of known errors separate the experimental data from
Eq.~(\ref{betatron}): electronic noise, uncertainties in the determination of
the lattice functions with errors, and BPM gain errors.

The first source of errors can be avoided if
average trajectories are used since this kind of trajectories have
very low noise levels. The second
source of errors has a small effect since the lattice functions with errors
are currently determined with an accuracy of
1\%~\cite{al_prst15,analyNBPM}. The third source of errors  can have a significantly larger effect on the experimental
data; it is the  dominant source of the three types of
errors.  Fortunately, large gain errors are not an
impediment to estimate accurately  $J_c$ and $\delta_c$. Because
Eqs.~(\ref{jandpsia})  and~(\ref{jandpsib})   allow  finding   a  value of  $J_c$ and
$\delta_c$  for every  pair of adjacent BPMs in the ring, a large
number of these measurements are available. If  the differences
between these measurements follow a Gaussian distribution,  the averages  values provide an  accurate measurement of $J_c$
and $\delta_c$ since their uncertainties should decrease as the
square root of the number of measurements.
\begin{figure}[h]
\centering
\includegraphics{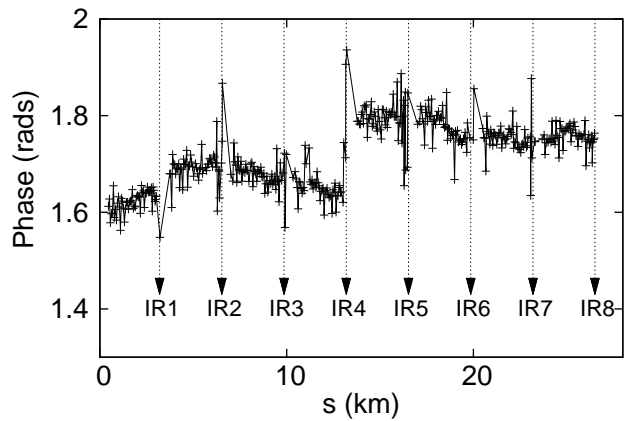}
\caption{\label{APJfig} Phase of  an average trajectory that was obtained using the 
  lattice functions with errors.  The
  average trajectory was built from an experimental TbT data set of beam 2. Phase measurements in or
  near  the IRs are not used since they have much larger
  fluctuations than phase measurements in the arcs.  The
  AC dipole is responsible for the jump that can be seen in IR4.}
\end{figure}

To estimate $J_c$ and $\delta_c$ from experimental data and evaluate
their accuracies,  action and phase plots are obtained  from  LHC
experimental TBT data  using the  lattice functions with errors.  It
can be seen that these plots are almost constant  for both beams and planes except for  jumps at the AC dipole location.
These jumps  are particularly strong in the $x$-plane of beam 2 (see
Fig.~\ref{APJfig}). Jumps in action and phase plots are due to
differences between the real model and the model that is actually
used to obtain these plots. The lattice functions used to obtain
Fig.~\ref{APJfig} do not include the effect of the  AC
dipole. Therefore, jumps are expected at this location.
\begin{figure}[h]
\centering
\includegraphics{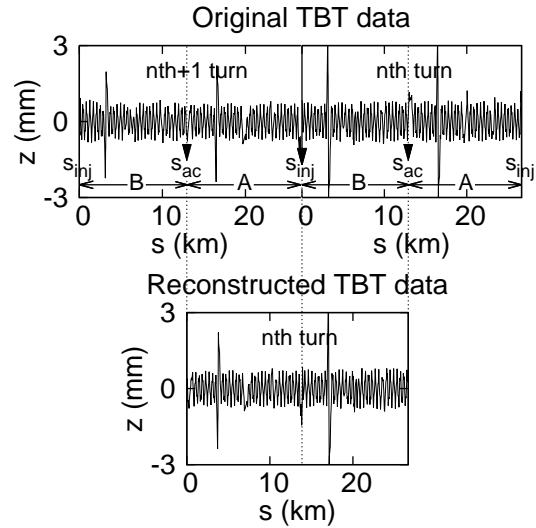}
\caption{\label{rotatelattice}  Reconstruction of the $n$th turn  from
  the $n$th +1   and $n$th turns of a  TBT data set. The $n$th turn of the
  reconstructed TBT data set is made of segment 
  ``A''  of the  original $n$th +1  turn and segment ``B''   of the original
  $n$th turn. In this way,  the reconstructed turn begins and ends at the
  longitudinal position of the AC dipole $s_{ac}$. The segments  ``A''  and  ``B''   are
  determined by  $s_{ac}$ and the longitudinal position the of
  injection point $s_{inj}$ as shown in the figure.} 
\end{figure}

\begin{figure}[h]
\centering
\includegraphics{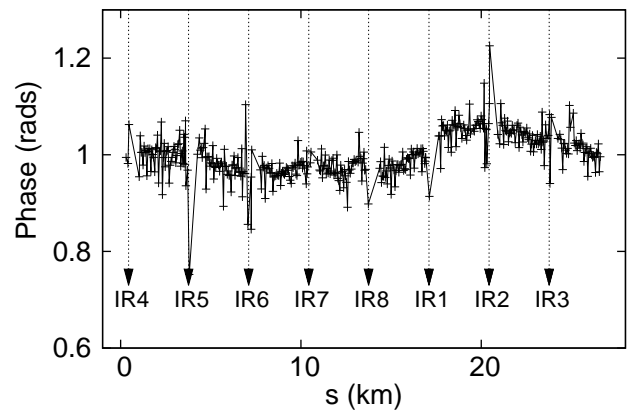}
\caption{\label{APJrot} Phase of an average trajectory obtained from 
  the new TbT data set  resulting from the reassignment of the longitudinal position of the original TBT data set used in
  Fig.~\ref{APJfig}. The longitudinal position of   $\beta_r(s)$ and $\psi_r(s)$ were also
  reassigned accordingly. Now it is possible to use all phase
  measurements (more than 400) to estimate $\delta_c$.}
\end{figure}

Having jumps in the middle of the action and phase plots is
not convenient since it limits the number of BPMs that can be used to
estimate $J_c$ and $\delta_c$. Fortunately,  the longitudinal position
originally assigned to the elements of  the accelerator lattice   can be
reassigned  so that the start  and
end points of the  action and phase plots correspond to the
location of the  AC dipole. In this way, the AC dipole jump  moves
toward the edge of the plots.

The reassignment of the longitudinal position must be performed for all
measurements and  functions  that are used to generate  the action and phase
plots, which are the  BPM measurements, $\beta_r(s)$, and $\psi_r(s)$.
To reassign the longitudinal position of  the BPMs measurements, every
turn of  a given TBT data set should be reconstructed as illustrated
in Fig.~\ref{rotatelattice}.

Reassignment of the longitudinal position of the  $\beta$-functions is done according to the difference between the injection
location and the AC dipole location. Reassignment of the longitudinal position  of the betatron phase functions
are similar to that of the  $\beta$-functions but, in addition,
the term $2\pi Q$  (with $Q$ the natural betatron tune) must be subtracted
from the original phases in the segment ``A''  (see
Fig.~\ref{rotatelattice}). After applying  APJ analysis to the same  TBT data set
used in Fig.~\ref{APJfig} and the same $\beta_r(s)$ and $\psi_r(s)$,
but with the longitudinal position reassigned, no significant jumps appear as can be seen in Fig.~\ref{APJrot}.

Now it is  possible to estimate $J_c$ and
$\delta_c$ with all  available  action and phase measurements as a
simple average.  The uncertainty associated with $J_c$ and $\delta_c$
is equal to the corresponding standard deviation  divided by the
square root of the number  measurements available in each case.  The
standard deviation for $\delta_c $ is not larger than 0.038 rads. Since the number of measurements is roughly 400, the  uncertainty
associated with $\delta_c$  is approximately 0.002 rads. The standard
deviation for $J_c$  is at most 10\%  of its average value,
so the corresponding uncertainty is less than 0.5\%.

\section{Four-quadrupole correction  from Experimental LHC Data}\label{appen}
The four-quadrupole correction was proposed and tested with only
simulations  in reference~\cite{jfc_prab17}. It is now possible to
estimate the  strengths for this kind of correction  from experimental
data  thanks to the new method that allows finding  $J_t$ and $\delta_t$ with very low uncertainties.

The strengths are mainly obtained from one-turn beam trajectories. To generate this kind of trajectories, a beam consisting of a
single bunch is excited transversally to  large amplitudes using
an AC dipole. This ensures that the beam  circulates for thousands of
turns without decoherence effects or  significant  growth of the bunch
size~\cite{acdipole}.  In every turn, all available BPMs  detect and measure the
transverse position of the  the bunch centroid, which results in
a one-turn beam trajectory. Since  the beam  circulates for thousands of
turns, thousands of one-turn trajectories are obtained every time the
AC dipole is activated. All these trajectories are saved in a file, which is
referred to as a TBT data set. In principle, only two one-turn trajectories are needed to make the
strength estimates. However,  the noise present in these
trajectories can generate  fluctuations in the corresponding action
and phases plots as large as the action and phase jumps used to
estimate the correction strengths. To solve this problem,
 special one-turn trajectories are built by selecting certain trajectories
 from the TBT data set and averaging them.  This results in what
 is called an average trajectory, which is finally the one-turn
 trajectory that is used to estimate correction strengths. The detailed procedure to build
 the average trajectories and the validity of using them  can be found  in~\cite{jfc_prab17}.

Before estimating the correction strengths, it is necessary to estimate the
quadrupole components of the  equivalent kicks from the average trajectories.  These estimates are made using Eqs.~(\ref{JIT}) to~(\ref{gammaR}), Eq.~(\ref{tetaexp}) and Eq.~(\ref{equierror}), which leads to  Table~\ref{quad_comp}.
\begin{table}[h]
\caption{ \label{quad_comp} Quadrupole components of the equivalent kicks due to magnetic errors in 
  the left and right triplets of IR1. All values given in units of $10^{-4}$ $m^{-1}$.}
\begin{ruledtabular}
\begin{tabular}{c D{,}{\pm}{-1}}
Left ${B_1}_{x,e}$&-9.70,0.04  \\
Left $\hat{B_1}_{y,e}$&-8.03,0.06  \\
Left $\hat{B_1}_{x,e}$&-4.95,0.05\\
Left ${B_1}_{y,e}$ &-7.15,0.05 \\
Right ${B_1}_{x,e}$&9.73,0.05  \\
Right $\hat{B_1}_{y,e}$&8.19,0.06 \\
Right $\hat{B_1}_{x,e}$ &7.91,0.04 \\
Right ${B_1}_{y,e}$ &9.93,0.06 \\
\end{tabular}
\end{ruledtabular}
\end{table}

The experimental data used  to obtain Table~\ref{quad_comp} consists of  five  TBT data sets of  beam 1, five
TBT data sets of beam 2,  and \textit{k}-modulation measurements
for both beams (Table~\ref{kmodexp}) that were taken in 2016.  For these experiments, IR1  was configured with a  nominal
$\beta^*$ of 40 cm, local and global coupling corrections were already
implemented, but normal quadrupoles corrections were off. To obtain the statistical uncertainties shown in
Table~\ref{quad_comp}, the   same procedure    was applied to every pair of TBT data sets  (one
TBT data set of beam 1 and one TBT data set of beam 2),  which resulted
in 5 different estimates for every quadrupole component.  The uncertainty  was
calculated as three times the standard deviation of these 5 estimates.

\begin{table}[h]
\caption{ \label{kmodexp} Values for the optical variables of the IR1
  inter-triplet space  obtained from \textit{k}-modulation
measurements.}
\begin{ruledtabular}
\begin{tabular}{cccc}
\multicolumn{1}{c}{\textrm{}}&
\multicolumn{1}{c}{\textrm{$\beta_r^*$}}&
\multicolumn{1}{c}{\textrm{$w_r$}}&
\multicolumn{1}{c}{\textrm{$\beta_{w_r} $}}\\
  &
\multicolumn{1}{c}{(cm)}   &
\multicolumn{1}{c}{(cm)}   &
\multicolumn{1}{c}{(cm)}   \\
\colrule
X - B1  & 86.1 & 43.0 & 40.7\\
Y - B1 & 70.3 & 33.8 & 44.9\\
X - B2  & 57.9 & 27.4& 38.2\\
Y - B2 &  70.0 & 35.2 & 39.7\\
\end{tabular}
\end{ruledtabular}
\end{table}

 Once the quadrupole components of the equivalent kicks are known, Eq.~(\ref{stcorr})  can be used to estimate the correction strengths from either beam 1 TBT data 
(method A) or from beam 2 TBT data (method B), which leads to Table~\ref{commcorrs_exp}.
\begin{table}[h]
\caption{\label{commcorrs_exp} Correction strengths obtained after
  applying Eq.~(\ref{stcorr}) on beam 1 experimental TBT data (method A) and beam 2 experimental TBT data (method B). The same
  procedure used in Table~\ref{quad_comp}  is used  to obtain the
statistical uncertainties shown.}
\begin{ruledtabular}
\begin{tabular}{c D{,}{\pm}{-1} D{,}{\pm}{6.4} }
& \multicolumn{2}{c}{Correction strengths}\\
& \multicolumn{2}{c}{($10^{-5}$  $m^{-2}$)}\\
Magnet &\multicolumn{1}{c}{A}
& \multicolumn{1}{c}{B}\\
\colrule
Q2L  &   1.28,0.01 &   1.20,0.02\\
Q2R &-0.97,0.02 &-0.70,0.01 \\
Q3L  &   1.83,0.02 &   1.07,0.03\\
Q3R  &-2.96, 0.04&-2.56,0.02\\
\end{tabular}
\end{ruledtabular}
\end{table}
The magnets used in the correction correspond to two quadrupoles of the left IR1 triplet (Q2L and Q3L) and two quadrupoles of the right IR1 triplet (Q2R and Q3R). Since these quadrupoles  are common to both beams, correction strengths obtained from  either beam 1 or beam 2 data should be identical. However, the
resulting  correction strengths are different for each case (columns A
and B of Table~\ref{commcorrs_exp}). These differences are
significantly larger than  the statistical uncertainties in
Table~\ref{commcorrs_exp}, specially for quadrupole Q3L. The presence
of magnetic errors in the matching quadrupoles can explain these
differences since these quadrupoles are no common to both beams. For this reason, a more general correction that takes into account  the matching quadrupoles was developed, and it is presented in the following section.    
\section{Corrections in the Matching Sections}\label{matching}
\begin{figure*}[t]
\centering
\includegraphics{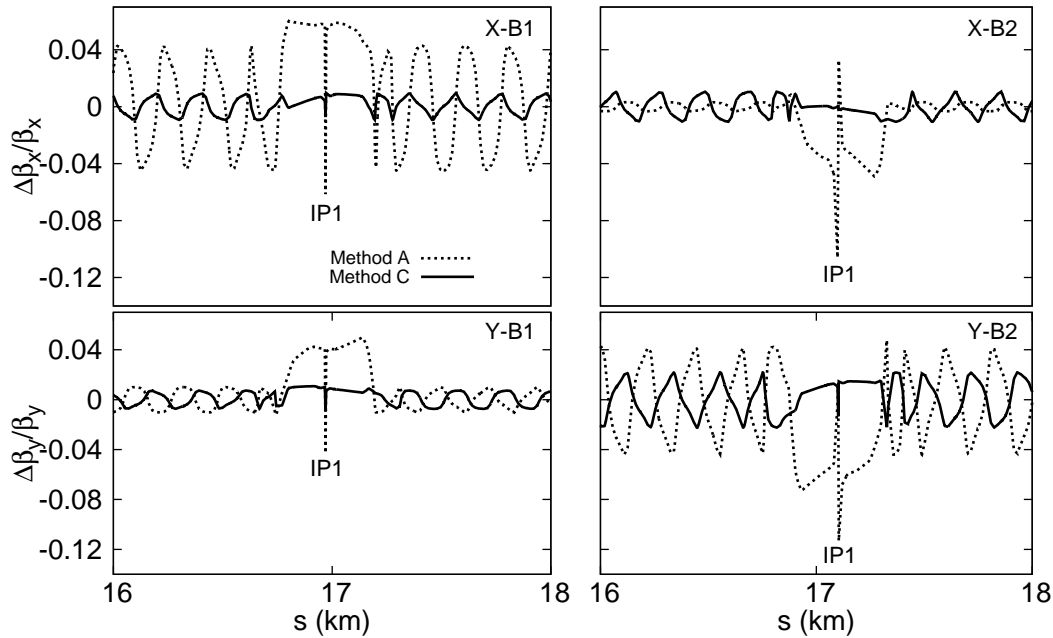}
\caption{\label{beatfig}  $\beta$-beating for the magnetic
  error distribution plus the corrections shown in
  Table~\ref{allcorrs_sim}. The residual $\beta$-beating after
  applying common corrector  are represented by dotted lines.  The residual $\beta$-beating  
 after applying  all correctors  are represented by solid lines. The
 dotted lines are obtained with correction strengths estimated with
 method A.  Similar results are obtained with method B.}
\end{figure*}

As mentioned earlier, the matching quadrupoles are located just
outside the triplets and there are two sets of these quadrupoles per
triplet (one per each beam), which leads to a total of 4 sets per
IR. Each set has magnets identified with labels Q4, Q5, Q6,
etc. In most cases, the matching quadrupoles   have betatron
phases that are very similar between them and their
corresponding triplet quadrupoles. Therefore, they are mathematically
equivalent to triplet quadrupoles and hence the same procedure used to
find Eq.~(\ref{intlincomp}) can be used to find expressions that
relate the quadrupole components of the equivalent kicks with their
correction strengths. Since the matching quadrupoles of  beam 1  are
independent of the matching quadrupoles of  beam 2, independent
expressions must be derived for each beam  leading to a system of four
equations. For the left IR1 side, those equations are
\begin{subequations}
 \label{allcorrs} 
\begin{eqnarray}
\hspace{-0.7cm} {B_1}_{x,e}  &=& -\frac{\Delta {K_1}_a I_{x,a}  + \Delta {K_1}_b  I_{x,b} +  \Delta
{K_1}_{c}  I_{x,c} }{\beta_{n,x}(s_e)}  \nonumber\\
\hspace{-0.7cm} &&  -\frac{\Delta {K_1}_{d}  I_{x,d}}{\beta_{n,x}(s_e)}, \qquad \qquad  \label{allcorrsa} \\
\hspace{-0.7cm}  {B_1}_{y,e}  &=&-\frac{\Delta {K_1}_a I_{y,a}
   + \Delta {K_1}_b  I_{y,b}  +  \Delta {K_1}_{c}
   I_{y,c}}{\beta_{n,y}(s_e)},   \label{allcorrsb}  \\
\hspace{-0.7cm} \hat{B_1}_{x,e}  &=& -\frac{\Delta {K_1}_a \hat{I}_{x,a}  + \Delta
  {K_1}_b  \hat{I}_{x,b} +  \hat{\Delta {K_1}_{c}}  \hat{I}_{x,c}
}{\hat{\beta}_{n,x}(s_e)},   \label{allcorrsc}  \\
\hspace{-0.7cm} \hat{B_1}_{y,e}  &=&-\frac{\Delta {K_1}_a \hat{I}_{y,a}
   + \Delta {K_1}_b  \hat{I}_{y,b}  +  \hat{\Delta
     {K_1}_{c}}\hat{I}_{y,c}}{\hat{\beta}_{n,y}(s_e)}    \nonumber \\
\hspace{-0.7cm} &&-\frac{\hat{\Delta {K_1}_{d}}  \hat{I}_{y,d}
}{\hat{\beta}_{n,y}(s_e)},  \label{allcorrsd}   
\end{eqnarray}
\end{subequations}
where $a$ and  $b$ correspond to the triplet
magnets Q2 and Q3, and $c$ and $d$ correspond to  the matching magnets
Q4 and Q6.  Q5 does not appear because its effect on the correction is equivalent to the
effect of Q4 except for a scale factor; therefore, only Q4 needs to be
activated. Other quadrupoles of the IR does not appear because their
beta functions are significantly lower than the beta functions of
quadrupoles Q1 to Q6. It should also be
noted that Q6  do not appear  in  Eqs.~(\ref{allcorrsb})
and~(\ref{allcorrsc}).  The corresponding terms have beta functions
and integrals that are very small and, therefore,  can be neglected.

The circumflex is used to distinguish the variables of beam 2 from those
corresponding to beam 1. Since magnets $a$ and $b$ are common to both
beams, no distinction should be made between beam 1  and beam 2 for the
correction strengths associated with these magnets.

There are 4 equations and 6 variables in Eq.~(\ref{allcorrs}); therefore, there
are infinite possible solutions.   A possible solution can be found if
the strengths of all no common correctors are initially forced to zero and the strengths
of  the common correctors are fitted to the resultant equations.  Once $\Delta {K_1}_a$ and  $\Delta
{K_1}_b$  are found, they can be substituted in the original set of
equations, and  a linear system of four-by-four  equations is obtained,
which can be solved by conventional methods.
 
The equations for the right side of IR1 are 
\begin{subequations}
 \label{allcorrs2} 
\begin{eqnarray}
\hspace{-0.7cm} {B_1}_{x,e}  &=& -\frac{\Delta {K_1}_a I_{x,a}  + \Delta {K_1}_b  I_{x,b} +  \Delta {K_1}_{c}  I_{x,c}}{\beta_{n,x}(s_e)}, \\
 \hspace{-0.7cm} {B_1}_{y,e}  &=&-\frac{\Delta {K_1}_a I_{y,a}
   + \Delta {K_1}_b  I_{y,b}  +  \Delta {K_1}_{c}  I_{y,c}
 }{\beta_{n,y}(s_e)} \nonumber \\
\hspace{-0.7cm} && -\frac{\Delta {K_1}_{d}  I_{y,d}}{\beta_{n,y}(s_e)}  ,    \\
\hspace{-0.7cm} \hat{B_1}_{x,e}  &=& -\frac{\Delta {K_1}_a \hat{I}_{x,a}  + \Delta {K_1}_b  \hat{I}_{x,b} +
  \hat{\Delta {K_1}_{c}}  \hat{I}_{x,c}}{\hat{\beta}_{n,x}(s_e)} \nonumber\\
\hspace{-0.7cm} &&-\frac{\hat{\Delta {K_1}_{d}}   \hat{I}_{x,d}}{\hat{\beta}_{n,x}(s_e)},   \\
 \hspace{-0.7cm}\hat{B_1}_{y,e}  &=&-\frac{\Delta {K_1}_a \hat{I}_{y,a}
   + \Delta {K_1}_b  \hat{I}_{y,b}  +  \hat{\Delta {K_1}_{c}}  \hat{I}_{y,c}}{\hat{\beta}_{n,y}(s_e)},   
\end{eqnarray}
\end{subequations}
which are solved following the same procedure employed for
Eq.~(\ref{allcorrs}).  After solving Eqs.~(\ref{allcorrs}) and~(\ref{allcorrs2}), a total 12
correction strengths can be found for the IR.

The validity of Eqs.~(\ref{allcorrs}) and~(\ref{allcorrs2}) can be
tested through simulated TBT data generated by  MADX in a LHC lattice
with a magnetic error distribution that includes magnetic errors in the matching quadrupoles. This magnetic error distribution  is created as
realistically as possible  (first column of Table~\ref{allcorrs_sim}).  For this purpose, the magnetic error distribution is chosen so that its equivalent quadrupole components
${B_1}_{z,e}$  are close to their
corresponding experimental values. This does not necessarily mean that
the magnetic error distribution corresponds to the actual error distribution. Due to degeneracy, there are infinite possible error distributions that reproduce the experimental ${B_1}_{z,e}$. 
\begin{table}[h]
\caption{ \label{allcorrs_sim} Strengths for a four-quadrupole
  correction estimated from beam 1 data (method A) and beam 2 data
  (method B). Also, the strengths for a twelve-quadrupole correction
  are shown in the last column (method C).  The
  suffixes B1 and B2 are used to distinguish the
  quadrupoles of beam 1 from the  quadrupoles of beam 2 }
\begin{ruledtabular}
\begin{tabular}{l D{.}{.}{2.2} D{.}{.}{2.2} D{.}{.}{2.2} D{.}{.}{2.2}}
& \multicolumn{1}{c}{Magnetic error}& \multicolumn{3}{c}{Correction strengths}\\
&\multicolumn{1}{c}{($10^{-5}$  $m^{-2}$)}& \multicolumn{3}{c}{($10^{-5}$  $m^{-2}$)}\\
\multicolumn{1}{c}{\textrm{Magnet}}& &
\multicolumn{1}{c}{\textrm{A}}&
\multicolumn{1}{c}{\textrm{B}}&
\multicolumn{1}{c}{\textrm{C}}\\
\colrule 
Q1L & -0.60 &\multicolumn{1}{c}{---} & \multicolumn{1}{c}{---} &\multicolumn{1}{c}{---}\\
Q1R  & 0.70 &\multicolumn{1}{c}{---} & \multicolumn{1}{c}{---} &\multicolumn{1}{c}{---}\\
Q2L &  -1.17 & 1.17& 1.00 & 1.08\\
Q2R  & 0.74 & -0.92& -0.62 & -0.77\\
Q3L  &  -1.31 & 1.90 & 1.21 & 1.55\\
Q3R & 2.60 & -2.97 & -2.62 & -2.79\\
Q4L.B1  & -7.00 & \multicolumn{1}{c}{---} & \multicolumn{1}{c}{---} & 10.92\\
Q4L.B2 & 7.00 &\multicolumn{1}{c}{---} & \multicolumn{1}{c}{---} & -10.94\\
Q4R.B1  &5.70&\multicolumn{1}{c}{---} &\multicolumn{1}{c}{---}  & -7.30\\
Q4R.B2 & -5.70& \multicolumn{1}{c}{---} &\multicolumn{1}{c}{---}  & 7.31\\
Q5L.B1  &  -6.86& \multicolumn{1}{c}{---} & \multicolumn{1}{c}{---} & \multicolumn{1}{c}{---}\\
Q5L.B2 &  7.01 & \multicolumn{1}{c}{---}& \multicolumn{1}{c}{---} & \multicolumn{1}{c}{---}\\
Q5R.B1  & 2.98& \multicolumn{1}{c}{---} & \multicolumn{1}{c}{---} &\multicolumn{1}{c}{---} \\
Q5R.B2& -3.45 & \multicolumn{1}{c}{---} & \multicolumn{1}{c}{---} &\multicolumn{1}{c}{---} \\
Q6L.B1  & 41.34 & \multicolumn{1}{c}{---} & \multicolumn{1}{c}{---} & -38.45\\
Q6L.B2 & -31.51 & \multicolumn{1}{c}{---}&\multicolumn{1}{c}{---}  & 32.02\\
Q6R.B1  &-23.71 &\multicolumn{1}{c}{---} & \multicolumn{1}{c}{---} & 22.05\\
Q6R.B2& 20.44 &\multicolumn{1}{c}{---} &\multicolumn{1}{c}{---} & -19.32\\
\end{tabular}
\end{ruledtabular}
\end{table}
Using the simulated TBT data generated with the error distribution
shown in the first column of Table~\ref{allcorrs_sim}, the strengths for a twelve-quadrupole correction
are estimated, which  results in  the last column (method C) of
Table~\ref{allcorrs_sim}.  The corresponding residual $\beta$ beating  (errors plus corrections), shown with the
solid lines in  Fig.~\ref{beatfig}, is below 4\% throughout the
ring, including the IP. The strengths of a four-quadrupole correction
are also estimated with data from beam 1 (method A) and beam 2 (method
B), leading to columns A and B of  Table~\ref{allcorrs_sim}.

Correction strengths obtained by  method A or method B can reduce
the $\beta$-beating to acceptable levels in the arcs, but the
$\beta$-beating in the IP can still be significant as shown by the
dotted lines in Fig.~\ref{beatfig}. With the method presented in this section (method C), the residual $\beta$-beating  in the arcs after applying this correction
is  smaller  than with  methods A and B, but more importantly,
the residual $\beta$-beating  at the IP is  significantly reduced.

The strengths of a six-quadrupole (all triplet quadrupoles) correction were also obtained
for the same  error distribution  using the SBS method \cite{hector_19}. The
corresponding residual $\beta$ beating is very similar to that
found with method A or B, that is, it is acceptable in the arcs but
very large in the IP.  
\section {Twelve-quadrupole correction  from experimental data and
  comparisons}\label{estimates}
The same experimental data used in Sec.~\ref{appen} is used in this section to obtain
the strengths of the twelve-quadrupole correction. The quadrupole
components of the equivalent kicks for these data  were already
estimated in that section and correspond to
Table~\ref{quad_comp}. Correction strengths are estimated by
applying the  procedure in the previous section to those quadrupole
components.The results are recorded in Table~\ref{allcorrstable}.

A comparison of the correction obtained in
Table~\ref{allcorrstable} can be made with a six-quadrupole correction
estimated with SBS, the method currently used in the LHC. In this
method, a variable related to the betatron phase called the phase error is defined. This variable is equal to
zero at some axial location just before the IR and starts to change as
a function of the axial coordinate due to the magnetic errors in the
IR.  Experimentally, the phase error is obtained from   Fourier
analysis of TBT data and the corresponding simulated phase  error is
derived from MADX simulations. To find the correction strengths, the six quadrupoles that participate in the correction are varied
iteratively  until the simulated phase error coincides with the
corresponding experimental  phase error. The variables  obtained from
k-modulation ($w_r$,  $\beta_{w_r}$  and $\beta^*$) and the $\beta$-beating in the IR also participate in this iteration process.
\begin{table}[h]
\caption{\label{allcorrstable} Correction strengths estimated for IR1
  from 2016 LHC data. The same
  procedure used in Table~\ref{quad_comp}  is used  to obtain the
statistical uncertainties shown.}
\begin{ruledtabular}
\begin{tabular}{l D{,}{\pm}{-1}}
 &
\multicolumn{1}{c}{Correction strengths}\\
Magnet
&
\multicolumn{1}{c}{($10^{-5}$  $m^{-2}$)}\\
\colrule
Q2L  &   1.24,0.01 \\
Q2R &-0.83,0.01  \\
Q3L  &   1.45,0.02 \\
Q3R  &-2.76, 0.02\\
Q4L.B1 &   11.0\mspace{9mu}, 0.35\\
Q4L.B2&-11.0\mspace{9mu},0.35\\
Q4R.B1 & -7.7\mspace{9mu},0.39 \\
Q4R.B2 &   7.7\mspace{9mu},0.39\\
Q6L.B1 &-40.1\mspace{9mu},1.3 \\
Q6L.B2 &   33.4\mspace{9mu},1.1 \\
Q6R.B1 &   24.2\mspace{9mu},1.4 \\
Q6R.B2 &-21.2\mspace{9mu}, 1.2\\
\end{tabular}
\end{ruledtabular}
\end{table}

 A comparison based simply 
on correction strengths may not be adequate due to degeneracy. A more reliable method to compare two different
correction should be based on their quadrupole components as explained
in Appendix \ref{degeneracy}. To make this comparison,  the correction strengths obtained in Table~\ref{allcorrstable} are directly substituted in  Eqs.~(\ref{allcorrs}), (\ref{allcorrs2}) and~(\ref{expcorrquads}) to obtain the first column
of Table~\ref{quad_corr}. As expected, the $B^{(c)}_{1z}$ are very
close to the corresponding ${B_1}_{z,e}$ obtained in Table~\ref{quad_comp} and also they have opposite signs so that they can cancel each other. The same procedure is
applied to the correction strengths obtained from the SBS method
(Table~II of~\cite{lhc_2015}) leading to the last column of
Table~\ref{quad_corr}.   An average  difference of about  20\%  can be observed between
the absolute values of the  quadrupole components of both
corrections.  An explanation of these
differences may be in the type of quadrupoles used in each correction.  SBS only uses triplet quadrupoles, while APJ also uses triplet and
matching quadrupoles.

\begin{table}[h]
\caption{\label{quad_corr} The quadrupole components of the equivalent
  kicks due to the correction strengths are estimated for two different methods of corrections. All values are given in units of $10^{-4}$ $m^{-1}$.}
\begin{ruledtabular}
\begin{tabular}{lcc }
\multicolumn{1}{c}{\textrm{}}&
\multicolumn{1}{c}{ APJ correction}&
\multicolumn{1}{c}{SBS correction}\\
\colrule
Left ${B}^{(c)}_{1x}$& 9.701  & 6.707 \\
Left $\hat{B}^{(c)}_{1y}$& 8.025  & 6.707 \\
Left $\hat{B}^{(c)}_{1x}$& 4.953  & 5.454 \\
Left ${B}^{(c)}_{1y}$  & 7.152 &  5.454 \\
Right ${B}^{(c)}_{1x}$& -9.726  & -6.055 \\
Right $\hat{B}^{(c)}_{1y}$& -8.186  & -6.055 \\
Right $\hat{B}^{(c)}_{1x}$ & -7.914  & -8.541 \\
Right ${B}^{(c)}_{1y}$ & -9.926&  -8.541 \\
\end{tabular}
\end{ruledtabular}
\end{table}

When only triplet quadrupoles are used for correction, the quadrupole
components of the equivalent kick due to   the correction strengths are
always subject to

\begin{align}
B^{(c)}_{1x}  & \approx  \hat{B}^{(c)}_{1y},  \\  
B^{(c)}_{1y} & \approx  \hat{B}^{(c)}_{1x},
\label{b1b2_relations} 
\end{align}
as  demonstrated in~\cite{jfc_prab17}. These symmetry relations
can be clearly seen in the quadrupole components generated by  the SBS
correction (last column of Table~\ref{quad_corr}). In contrast, the
experimental quadrupole components do not show these symmetries and,
therefore,  can not be  completely  compensated with a correction
that only uses triplet quadrupoles.

SBS corrections during the 2016 LHC were considered essential to
achieve  a rms $\beta$-beating below 2\% around the rings. But
achieving this  low $\beta$-beating does not necessarily imply that
SBS corrections perfectly compensate the magnetic errors in the
IRs; there is also the possibility that the global correction that was
applied  later through the matching and dispersion suppressor quadrupole  also compensated the
 residual $\beta$-beating left by SBS corrections in every IR.

\section{Conclusions}
 Mathematical relationships that allow estimating the action and phase  in the 
inter-triplet space were deduced. It was shown that
the  uncertainties associated with these formulas were significantly
lower than the uncertainties of  a method   that uses two
BPMs in the inter-triplet space. This last  method required a very precise calibration of the
BPMs, which is not yet available in the LHC. In contrast,  the new
method can be used to make reliable estimates of action and
phase in the inter-triplet space  with the hardware currently  available in the LHC. 

Strengths of a  four-quadrupole correction for IR1 were estimated from
experimental LHC data. These strengths were estimated independently
for each beam giving different values, suggesting that magnetic
errors in the no common quadrupoles of the IR  were significant. As a consequence,
a more general correction scheme that uses twelve quadrupoles was
developed and tested with simulations. These simulations show that
the twelve-quadrupole correction can suppress the $\beta$ beating
generated in the IR throughout the ring, including the IP, even when
there are large magnetic errors in the matching quadrupoles. In
contrast, the four-quadrupole and six-quadrupole correction, either
estimated with APJ or  SBS, cannot  guarantee suppression of the
$\beta$ beating in the IP under these conditions.
 
The  strengths of a twelve-quadrupole correction  in IR1 were also
estimated from LHC experimental data. The resulting correction was
compared to a correction obtained in similar conditions with the SBS
method.  The comparison was made through the quadrupole components
associated with the corrections obtained with  each method.  An
average difference  of 20\% was found between the quadrupole
components  associated with each method. The fact that the SBS correction does not use the matching quadrupole as
correctors, only the triplet quadrupoles, probably explains these
differences. The SBS correction
is acceptable if the residual $\beta$-beating is subsequently
suppressed through global corrections  as it is currently done in the
LHC.  If full local compensation  is required, the  matching quadrupole should be included in
the correction as proposed in this paper.

The IR corrections in the LHC Run 3 in 2021 are expected to be
significantly different to the corrections found during the LHC Run 2 or
the LHC Run 1. The  method presented in this paper is a viable option  to recalculate those corrections.
\section*{Acknowledgments}
We are very thankful to all members of  the optics measurement and correction team
(OMC) at CERN for support with their  \textit{k}-modulation  software,
GetLLM  program, and  experimental data. Special thanks goes to Hector Garc\'ia Morales, member of the OMC team, for analysis of simulations related with the SBS correction method. Y. Rodr\'iguez wants to
thank the support received through  ``CONVOCATORIA NACIONAL PARA EL APOYO  A LA MOVILIDAD INTERNACIONAL DE
LA UNIVERSIDAD NACIONAL DE COLOMBIA 2017-2018'', which made possible
a short stay at CERN. 
\appendix
\section{Degeneracy in the Corrections}\label{degeneracy}
The quadrupole strengths of at least two quadrupoles in a triplet must
be changed to suppress the quadrupole component of the equivalent kick
associated with that triplet,  as indicated in
Sec.~\ref{apjsection}. This suppression can also be done by changing
the strengths of the three quadrupoles.  In that case,
Eqs.~(\ref{expcorrquads}) and~(\ref{intlincomp}) becomes \par
\begin{align}
B^{(c)}_{1x}  &=& \frac{1}{\beta_{n,x}(s_e)}
\left( \Delta {K_1}_a I_{x,a}  + \Delta {K_1}_b  I_{x,b}   +  \Delta
  {K_1}_c  I_{x,c}  \right) \nonumber\\
&&\\
 B^{(c)}_{1y} &=&\frac{1}{\beta_{n,y}(s_e)} \left( \Delta {K_1}_a I_{y,a}  +
   \Delta {K_1}_b  I_{y,b} + \Delta {K_1}_c  I_{y,c}  \right).\nonumber
\end{align}
Since there are 3 quadrupoles whose strengths can be changed  and only two
equations, there are infinite possible ways of generating the same
quadrupole components.  On the other hand, the $\beta$-beating generated by the correctors in
the triplets is given by 

\begin{align}
\label{betaandquads}
\left(\frac{\Delta \beta_x(s)}{\beta_x(s)} \right)_{TR} &=&
-\frac{\cos\left[ 2 |\psi_{n,x}(s) -\psi_{n,x}(s_e)|  -2 \pi
    Q_x\right]}{2\sin2\pi Q_x} \nonumber\\
  &&\times \beta_{n,x}(s_e)   B^{(c)}_{1x},  \nonumber\\
&&\\
\left(\frac{\Delta \beta_y(s)}{\beta_y(s)} \right)_{TR} &=&
-\frac{\cos\left[ 2 |\psi_{n,y}(s) -\psi_{n,y}(s_e)|  -2 \pi
    Q_y\right]}{2\sin2\pi Q_y} \nonumber\\
  &&\times \beta_{n,y}(s_e)   B^{(c)}_{1y},  \nonumber
\end{align}
where Eq.~(D4) of~\cite{jfc_prab17}  was used. Since the $\beta$-beatings are
proportional to the quadrupole components,
there are infinite sets of correction strengths that generate the same
$\beta$-beatings. This means that corrections with very different correction
strengths can have the same effect in the accelerator
optics. Therefore, if a comparison between two different corrections is
required, it should be done by comparing their quadrupole components
instead of their individual correction strengths. This demonstration
can also be extended for the case in which the matching quadrupoles are
also used in the correction.

%
\end{document}